\documentstyle[12pt,epsfig,amsmath,amssymb]{article}

\newcommand{\beq}{\begin{equation}}
\newcommand{\eeq}{\end{equation}}
\newcommand{\bq}{\begin{equation}}
\newcommand{\eq}{\end{equation}}
\newcommand{\ba}{\begin{array}}
\newcommand{\ea}{\end{array}}
\newcommand{\beqa}{\begin{eqnarray}}
\newcommand{\eeqa}{\end{eqnarray}}
\def\bc{\begin{center}}
\def\ec{\end{center}}
\def\bnum{\begin{enumerate} }
\def\enum{\end{enumerate}}
\def\nn{\nonumber}

\def\leqn#1{(\ref{#1})}

\def\nn{\nonumber}

\def\[{\left[}
\def\]{\right]}
\def\({\left(}
\def\){\right)}

\def\>{\rightarrow}

\def\vareps{\varepsilon}

\def\Dslash{\not{\hbox{\kern-4pt $D$}}}
\def\pslash{\not{\hbox{\kern-4pt $p$}}}
\def\qslash{\not{\hbox{\kern-4pt $q$}}}
\def\lv{\not{\hbox{\kern-4pt $L$}}}
\def\lsim{\mathrel{\raise.3ex\hbox{$<$\kern-.75em\lower1ex\hbox{$\sim$}}}}
\def\gsim{\mathrel{\raise.3ex\hbox{$>$\kern-.75em\lower1ex\hbox{$\sim$}}}}
\def\ifmath#1{\relax\ifmmode #1\else $#1$\fi}

\evensidemargin=\oddsidemargin

\begin{document}

\begin{titlepage}
\begin{flushright}
\end{flushright}

\begin{center}
 \vspace*{10mm}

{\huge\bf Topological Interactions in the Higgsless \\ Model at the LHC}\\

\medskip
\bigskip\vspace{0.6cm}
{
{\large\bf Maxim Perelstein$^{\dag}$, Yong-Hui Qi$^{\dag,\star}$}
}
\\[7mm]
{\it $^{\dag}$ Institute for High Energy Phenomenology,\\ Newman Laboratory of Elementary Particle Physics,\\
Cornell University, Ithaca, NY 14853, USA\\}
\vspace*{0.3cm}
{\it $^{\star}$ Physics Department,\\
Center for High Energy Physics,\\
Tsinghua University, Beijing, 100086, China}
\\
\vspace*{0.3cm}
{\tt  mp325@cornell.edu},\quad {\tt yq32@cornell.edu}
\bigskip\bigskip\bigskip

{
\centerline{\large\bf Abstract}
\begin{quote}
Topological quantum interactions, namely Chern-Simons terms and global Wess-Zumino terms, arise naturally in
theories with extra dimensions of space compactified on orbifolds. If the extra dimensions become manifest at the TeV scale, experiments at the Large Hadron Collider (LHC) could observe signatures of topological
interactions. Decays of Kaluza-Klein excitations of neutral electroweak gauge bosons into
pairs of neutral Standard Model gauge bosons, $Z^0Z^0$ and $Z^0\gamma$, would provide a clean
signature, since such decays do not occur at tree level. In
this paper, we investigate the prospects for discovering such decays at the LHC, in the context of the Higgsless model of electroweak symmetry breaking. We identify the form of the relevant topological interactions, and estimate their strength. We find that in the minimal version of the model, the signal
may be observed with about 100 fb$^{-1}$ of data at the LHC using the Drell-Yan production process of the Kaluza-Klein gauge bosons. In addition, it is likely that the ultraviolet completion of the model would contain additional massive fermions, which can significantly enhance the signal. With two additional fermion multiplets, observation of the topological decay modes at the 3-sigma level would be possible with about 100 fb$^{-1}$ of data using the highly model-independent vector boson fusion production channel.
\\
\end{quote}}
\end{center}
\end{titlepage}

\section{Introduction}

In spite of spectacular successes of the Standard Model (SM) of high-energy physics, the mechanism of electroweak symmetry breaking (EWSB) is still unknown. Many candidate theories of this phenomenon have been proposed and studied. These theories can be divided into two broad categories, those with an elementary Higgs (e.g. supersymmetry), and those with a composite Higgs or with no Higgs at all (e.g. technicolor). Historically, theoretical analysis of models of the second class has been impeded by the fact that they generically become non-perturbative at energies within a factor of ten above the EWSB scale, i.e. close to the TeV. More recently, a number of interesting new models in this class have been proposed, which can be at least partially analyzed within perturbation theory. These models extend the usual space-time by adding an extra compact spatial dimension, typically with the geometry of a slice of anti-de Sitter (AdS) space. Gauge theories propagating on this space are conjectured to be dual to four-dimensional (4D), strongly-coupled conformal field theories (CFTs), with the conformal symmetry broken in the infrared~\cite{duality}. (While the details of this correspondence are at present not precisely understood, it is widely expected to hold, given the extensive evidence for similar dualities in supersymmetric theories~\cite{SUSYduality}.)
As long as the 5-dimensional (5D) theory remains weakly coupled, its physics can be studied using perturbative techniques. Examples of 5D EWSB models constructed using this approach include the original Randall-Sundrum model~\cite{RS}, the 5D composite Higgs model~\cite{comp}, and the Higgsless model~\cite{HL}.

It is well known that the 5D gauge Lagrangian admits the Chern-Simons (CS) term, a renormalizable operator with no equivalent in 4D. This term is gauge-invariant everywhere {\it except} at the boundaries of the 5D space. In the 5D EWSB models of interest, the boundaries are two 4D hypersurfaces at the edges of the AdS slice. Since Einstein equations demand that non-zero energy density be localized on these surfaces, it is natural to assume that physical objects (e.g. D3-branes of string theory) are present there. 4D fermions can be localized on these branes. If these fermions are chiral, they may induce gauge anomalies. As long as the low-energy 4D theory is anomaly-free (which is required in any realistic model), the anomalies on the two branes can be simultaneously
canceled by the CS term, restoring gauge invariance at every point in the 5D theory. Thus, for a given choice of the boundary fermion content, the coefficient of the CS term is uniquely fixed by the requirement of gauge invariance in 5D.

CS terms also naturally arise in theories where all fermions are free to propagate in the full 5D space (the ``bulk"). In such theories, each 5D fermion can be viewed as an infinite tower of 4D fields, Kaluza-Klein (KK) modes. With an appropriate choice of boundary conditions on the two branes, the lowest KK mode (the ``zero-mode") is a massless Weyl fermion, which can be identified with an SM fermion. The existence of the zero-mode is independent of the 5D mass of the original fermion field, but the wavefunction (or profile) of this zero-mode in extra dimension depends sensitively on this parameter. In many cases, the zero-mode is quasi-localized on one of the branes, {\it i.e.} the wavefunction is exponentially enhanced close to a brane. It is reasonable to model such zero modes as brane-localized 4D fermions. These may induce anomalies on branes, requiring a CS term to cancel them. In this case, the CS term does not need to be added by hand, but instead arises with exactly the required coefficient when the excited KK modes of the 5D fermions are integrated out~\cite{Nima,Skiba}.

If present, CS terms in the 5D EWSB models can have interesting observable consequences. For example, their presence may lead to explicit violation of Kaluza-Klein parity~\cite{HillNoKK}. In models where the KK parity is imposed at tree level, the CS couplings provide the dominant decay mode of the lightest of the would-be KK-odd particles, giving spectacular multi-gauge-boson events at the LHC~\cite{NoKKpheno}.
In models where the KK parity is {\it not} imposed at tree level (such as the Randall-Sundrum and Higgsless models) the consequences of the CS terms are less dramatic, but may still be observable.
Some of them have been discussed recently by Bai, Burdman and Hill~\cite{BBH}, who analyzed, in particular, the prospects for observation of the associated production of first-level and second-level KK gluons at the LHC.
In this paper, we will study another novel aspect of collider phenomenology due to the CS terms: Decays of the KK-excited neutral electroweak gauge bosons into $ZZ$ and $Z\gamma$ pairs.\footnote{Note that $ZZ$ and $Z\gamma$ decays of a techni-omega particle of technicolor theories have been considered in the past, {\it e.g.} in Ref.~\cite{CG}. (See also Ref.~\cite{TCreview} for a review and further references, and Ref.~\cite{LH2008} for a recent LHC study.)  Another related recent study in a different context appears in Ref.~\cite{James}.} Such decays are not induced at the tree level by the usual gauge interactions, since all three bosons are neutral. Thus, these channels are a good place to search for small effects, such as the CS terms.\footnote{Topological terms are not the only type of new physics that could induce such decays. For example, in the two-Higgs-doublet model, the decay $Z^\prime\to ZZ$ can be induced at the one-loop level in the presence of CP violation~\cite{ref10}. These two models could in principle be distinguished experimentally, as explained in Ref.~\cite{Ian}.}
For concreteness, we will perform our study in the context of the semi-realistic ``cured" 5D Higgsless model of Ref.~\cite{HL_S}. In this model, the $S$ and $T$ parameters are consistent with precision electroweak fits, while at the same time KK-excited gauge bosons  have masses in the range 0.6-1.2 TeV, and can be produced at the LHC with fairly large cross sections. The ``cured" model is not fully realistic because the predicted correction to the $Zb\bar{b}$ coupling is too large. This can be addressed by expanding the 5D space further and adding another brane to lift the top mass~\cite{2branes}, or by incorporating the custodial symmetry of Ref.~\cite{cusZbbar} to protect the $Zb\bar{b}$ coupling~\cite{HL_cust}. The resulting models are somewhat more complicated, and we will not consider them here; however, we expect that our main conclusions apply, at a qualitative level, to these versions of the model.

The rest of the paper is organized as follows: In Section~\ref{sec:model} we outline the main features of the ``cured" Higgsless model. We then discuss the Chern-Simons terms, and the closely related ``global Wess-Zumino" (WZ) interactions, present in this model, and estimate their coefficients. We discuss the decays of neutral electroweak KK gauge bosons, both conventional (induced by Yukawa and gauge couplings) and anomalous (induced by the CS and WZ terms) in Section~\ref{sec:decays}. In Section~\ref{sec:LHC}, we discuss collider signatures of the anomalous decays, estimate the relevant backgrounds, and comment on the prospects for observing these modes at the LHC. Our conclusions, and some possible directions for future work, are presented in Section~\ref{sec:concl}.

\section{The Model}
\label{sec:model}

In 5D Higgsless models, the $W$ and $Z$ bosons get their masses through appropriate boundary conditions on 5D gauge fields, and the EWSB scale is effectively fixed by the size of the extra dimension. Unitarity violation in gauge bosons scattering amplitudes can be delayed via heavy KK modes exchange~\cite{Csaki:2003dt}, so that the TeV-scale physics in these models can be analyzed within perturbation theory. In this study, we will use the ``cured" Higgsless model on a warped gravitational background proposed in Ref.~\cite{HL_S}. This model satisfies most electroweak precision tests, and predicts the first set of KK excitations of electroweak gauge bosons in the 600 GeV-1.2 TeV range, which can be copiously produced and studied by the LHC. After briefly reviewing the model, we will discuss the form of the Chern-Simons terms and the related global WZ interactions, and estimate their coefficients.

\subsection{The ``Cured" Higgsless Model}
\label{sec:min}

We work on the $AdS_{5}$ background and parameterize the space-time using conformal coordinates
\beq
ds^{2}=\left(\frac{R}{z}\right)^{2}(\eta_{\mu\nu}dx^{\mu}dx^{\nu}-dz^{2})\,.
\eeq
The fifth dimension is an interval $z\in [R^\prime, R]$. The branes are located at the endpoints of the interval, with the ``ultraviolet (UV) brane" at $R$ and the ``infrared (IR) brane" at $R^\prime$.
A 5D gauge theory lives on this space, with the gauge group $SU(3)_{c}\times SU(2)_{L}\times SU(2)_{R}\times U(1)_{X}$. (In the minimal model, $X=(B-L)/2$.) The bulk gauge group is broken by boundary conditions to $SU(3)_c\times SU(2)_{L}\times U(1)_{Y}$ on the UV brane and $SU(3)_c\times SU(2)_{D}\times U(1)_{X}$ on the IR brane, where $SU(2)_D$ is the diagonal combination of the two $SU(2)$ factors. We assume that all brane-localized gauge kinetic terms are negligible. The gauge sector of the theory then is parametrized by four 5D gauge couplings,
in addition to the scales $R$ and $R^\prime$. For simplicity, we will assume throughout this paper that the 5D couplings of the $SU(2)_L$ and $SU(2)_R$ factors are equal, for a total of 5 free parameters. Reproducing the SM in the low-energy limit fixes four combinations of these parameters, the three 4D gauge couplings and the $Z$ mass. (The $W$ mass automatically satisfies the SM relation $M_W=M_Z\cos\theta_w$ at the tree level, since the $SU(2)_R$ acts as the custodial symmetry in 4D~\cite{Raman_cust}.) The gauge sector of the model then has one free parameter. It is convenient to use the mass of the first KK-excited neutral electroweak gauge boson to parameterize this freedom.

SM fermions are incorporated as zero-modes of 5D fields. Our notation for a 5D fermion is
\beq
\Psi=\left(
       \begin{array}{c}
         \chi \\
         \bar{\psi} \\
       \end{array}
     \right)\,,
\eeq
where $\chi$ transforms as a left-handed Weyl fermion under the 4D Lorentz group, and $\psi$ transforms as a right-handed Weyl fermion. The simplest embedding of one generation of quarks is:
\beq
  \begin{array}{cccc}
      & SU(2)_{L} & SU(2)_{R} & U(1)_{X} \\
    Q_{L}=\left(
            \begin{array}{c}
              u \\
              d \\
            \end{array}
          \right)_{L}  & \square & 1 & 1/6 \\
    Q_{R}=\left(
      \begin{array}{c}
        u \\
        d \\
      \end{array}
    \right)_{R}
     & 1 & \square & 1/6 \\
  \end{array}\label{Rep2generations}
\eeq
where $u_L$, $u_R$, $d_L$ and $d_R$ are 5D fermion fields.
(The lepton sector is identical, but with $X=-\frac{1}{2}$.) The boundary conditions for the up-type quark are
\beq
\psi_{u_{L}}|_{R,R^{'}}=0\qquad \chi_{u_{R}}|_{R,R^{'}}=0\,.
\eeq
This yields a left-handed zero mode for $u_{L}$ and a right-handed zero mode for $u_{R}$. These zero-modes are identified with the SM up-quark.
The same boundary conditions are chosen for down-type quarks. On the IR brane, both $Q_{L}$ and $Q_{R}$ are $SU(2)_{D}$ doublets, and a Dirac mass term is allowed:
\beq
S_{IR}=\int d^{5}x\,\left(\frac{R}{z}\right)^{4}\,\delta(z-R^{'})M_{D}R^{'}\bar{Q}_{L}Q_{R}+h.c.
\eeq
This term modifies the boundary conditions on the IR brane, and provides the SM fermions with their masses.

The model allows for a 5D ``bulk" mass for each fermion. The existence and chirality of the zero modes is determined solely by boundary conditions, and is independent of  the 5D masses. However, the wavefunction of the zero mode in the 5th dimension is affected by the bulk mass: it is proportional to $(\frac{z}{R})^{2-c_L}$ for left-handed zero modes and $(\frac{z}{R})^{2+c_R}$ for right-handed zero modes~\cite{fermion_profile}, where $c_{L,R}$ is the bulk mass in units of the bulk curvature $k=1/R$. As a result, the fermion zero mode couplings to KK-excited gauge bosons, which are proportional to the overlap integral of their wavefunctions, are $c$-dependent. Exchanges of KK-excited gauge bosons in SM fermion scattering contribute to precision electroweak observables, and to satisfy precision electroweak constraints (in particular the $S$-parameter bound) these contributions need to be suppressed. This can be achieved by choosing the fermion zero mode wavefunctions to be (approximately) orthogonal to the KK gauge boson profiles. It has been shown~\cite{HL_S} that this occurs for $c_L \approx -c_R \approx 0.5$, that is, ``ideally delocalized" fermions with flat profiles in the fifth dimension. We will assume $c_L=-c_R=0.46$ for all leptons, and for the first two generations of quarks. This choice does not allow for sufficiently large top quark mass, which requires that some of the third-generation quark zero modes be localized near the IR brane. We will assume $c=0.46$ for $Q_L^3$, and $c=-0.05$ for $Q_R^3$~\cite{HL_S}. Once the bulk masses are fixed, the IR brane Dirac masses are chosen to reproduce the SM fermion spectrum. We assume that all brane-localized fermion kinetic terms are negligible. With these choices, no new parametric freedom is introduced in the fermion sector, so we remain within a one-parameter model.

The above construction will be referred to as the ``minimal model" in our study. We will also consider a
simple extension with the same zero-mode spectrum but additional massive fermions at the TeV scale, which can contribute to the topological interactions. This ``non-minimal model" is defined in Sec.~\ref{sec:non-min} below.

\subsection{Chern-Simons Terms}

A general non-Abelian Chern-Simons (CS) term in 5D has the form
\beq
S_{CS}=c_{CS} \int d^5x {\rm Tr}\,\left[A dA dA +\frac{3}{2}A^{3}dA +\frac{3}{5}A^5\right]\,,\label{CS5}
\eeq
where we used differential form notation: the Yang-Mills gauge field 1-form is
\beq
A=-iA_M^{a}T^{a}dx^M
\eeq
and the field strength is a 2-form
\beq
F=dA+A^2=-i\frac{1}{2}F_{MN}^{a}T^{a}dx^{M}\wedge dx^{N}\,.
\eeq
Here $T^a$ are the gauge group generators normalized by Tr$(T^aT^b)=\frac{1}{2}\delta^{ab}$.
The purely-abelian CS Lagrangian has the same form, but the last two terms vanish due to $A^2=0$. In a theory with a product gauge group, such as the Higgsless model, a variety of CS terms are possible.
The CS terms are gauge-invariant in the bulk, but have non-trivial gauge variation on the boundaries. Thus, in a gauge-invariant theory on an interval, CS terms can only be present if the gauge symmetry is anomalous on the boundaries, for example due to chiral fermions localized on the branes. The requirement of gauge invariance then fixes the coefficient of the CS term. (Note that the gauge anomalies on the two branes must be equal and opposite for this to work; this is of course also the condition for the low-energy 4D theory to be anomaly-free, and so is always satisfied in models we consider.)

Using dimensional deconstruction approach, it has been shown in Ref.~\cite{BBH} that
if a bulk fermion has a zero-mode  localized near one of the branes, it can be replaced, for the purposes of computing the coefficient of the CS term, with a 4D fermion on that brane. If the zero-modes are delocalized, and the left and right-handed zero-modes have the same profile, as is the case for the first two generations of our model, 4D anomalies cancel point-by-point in the bulk, and no CS term is needed. For the third generation, we will make the following simple approximation: we assume that $Q^3_L$, $L^3_L$, and $L^3_R$ zero-modes are localized on the UV brane, while $Q^3_R$ zero-mode is localized on the IR brane. This is the same approximation as was made in Ref.~\cite{BBH}, and should capture the physics reasonably well, although some corrections from less-than-ideal localization in the realistic Higgsless model are inevitable. (We defer computing these corrections to future work.) Note that this assumption is {\it only} made in the calculation of the CS coefficients; for the rest of the analysis, such as computing couplings of gauge KK modes to SM fermions, we use exact zero-mode profiles.

Within this approximation, anomaly cancellation for the 3rd generation works as follows:
$SU(2)_L^2\times U(1)$ anomaly cancels locally on the UV brane, and does not require a CS term. All other gauge anomalies -- $SU(3)^3$, $SU(3)^2\times U(1)$, $SU(2)_R^2\times U(1)$, and $U(1)^3$ -- are not cancelled on each brane, and CS terms are required for gauge invariance. Since we focus on electroweak sector here, we are only interested in the last two anomalies. Moreover, we will restrict our attention to terms cubic in the gauge fields, since they are the only ones contributing to the phenomenological signatures we will study. The corresponding CS terms are
\beqa
S_{CS5}&=&\frac{1}{24\pi^2}\,\int d^4x\int dz \,\epsilon^{\mu\nu\rho\sigma}\Bigl[  \,\frac{1}{72}\, (\partial_{z}B_{\mu}B_{\nu}-B_{\mu}\partial_{z}B_{\nu})\, B_{\rho\sigma}\nn\\
 \,&+& \,
\frac{1}{8} \,\left(\partial_{z}B_{\mu}A_{R\nu}^{3}-A_{R\mu}^{3}\partial_{z}B_{\nu}-B_{\mu}\partial_{z}A_{R\nu}^{3}+\partial_{z}A_{R\mu}^{3}B_{\nu}\right)\,F_{R\rho\sigma}^{3}\,\Bigr]\,,
\label{CS}
\eeqa
where we used the unitary gauge ($A_5=0$) for all gauge fields, and Greek indices run over the 4 non-compact dimensions. Here $B$ is the $U(1)_X$ gauge field and $A_R$ is the $SU(2)_R$ gauge field.

The CS terms can be thought of as the result of integrating out KK-excited modes of the bulk fermions. The main object of interest to us in this paper are the vertices involving three gauge bosons. In addition to a contribution from the CS terms, these vertices receive contributions from the one-loop triangle diagrams with {\it zero-mode} fermions running in the loop. This contribution can be expressed as a ``global Wess-Zumino" effective action~\cite{Skiba,BBH}, which has the form (in unitary gauge)
\beq
\Gamma_{WZ}=-c\int d^4x \, [B(R^{'})dB(R^{'})B(R)-B(R)dB(R)B(R^{'})]
\eeq
for an Abelian field, and
\beqa
&\Gamma_{WZ}=-c\int d^4x \, \text{Tr}\{\frac{1}{2}[A(R^{'})dA(R^{'})A(R)-A(R)dA(R)A(R^{'})
+dA(R^{'})A(R^{'})A(R)\nn\\ & -dA(R)A(R)A(R^{'})]+\frac{1}{2}(A(R^{'})^{3}A(R)-A(R)^{3}A(R^{'})-\frac{1}{2}A(R)A(R^{'})A(R)A(R^{'}))\}
\eeqa
for a non-Abelian field. The factors $c$ are proportional to the corresponding anomaly coefficients.
In our model, the WZ effective action in the electroweak sector has the form
\beqa
&\Gamma_{WZ}=-\frac{1}{24\pi^{2}}\int d^4x \,\epsilon^{\mu\nu\rho\sigma} \,\frac{1}{144}\left[B_{\mu}(R^{\prime})B_{\nu}(R)-B_{\mu}(R)B_{\nu}(R^{\prime})\right]\,\times\,\left[B_{\rho\sigma}(R^{'})+B_{\rho\sigma}(R)\right]\nn\\
&-\frac{1}{24\pi^{2}}\int d^4x \, \epsilon^{\mu\nu\rho\sigma} \,\frac{1}{16}\,\left[\,B_{\mu}(R^{\prime})A_{R{\nu}}^{3}(R)-B_{\mu}(R)A_{R{\nu}}^{3}(R^{\prime})-A_{R{\mu}}^{3}(R)B_{\nu}(R^{'})+A_{R{\mu}}^{3}(R^{'})B_{\nu}(R)\,\right]\nn\\
&\times \, \left[F_{R\rho\sigma}^{3}(R)+F_{R\rho\sigma}^{3}(R^{'})\right]\,.
\label{WZ}
\eeqa
Combining Eqs.~\leqn{CS} and~\leqn{WZ} yields the effective ``topological" Lagrangian for anomalous triple-gauge-boson interactions in the electroweak sector of our model. Our task is to investigate a collider signature of these interactions.

\subsection{Non-Minimal Model}
\label{sec:non-min}

In addition to the fermion fields whose zero modes are required to reproduce the SM particle content,
the Higgsless model may also contain other fermions, with masses too high to be discovered up to now. While in 4D such heavy fermions are typically vector-like and do not contribute to anomalies, in 5D constructions this does not need to be the case. For example, imagine adding a pair of 4D Weyl fermions:
$X_L$ localized on the UV brane and $X_R$ on the IR brane, transforming in the following representations of $SU(3) \times SU(2)_{L} \times SU(2)_{R} \times U(1)_{X}$:
$X_{R}\,\in\, ({\bf N}_c, {\bf N}_L, {\bf N}_R, x)$; $X_{L}\,\in\, (\overline{{\bf N}}_c, \overline{{\bf N}}_L, \overline{{\bf N}}_R, -x)$. While the two form a vector-like pair from the 4D point of view, their spatial separation in 5D means that a CS term needs to be induced to locally cancel anomalies on the branes. A mass term with a 5D Wilson link, of the form
\beq
L_{m} \,=\, M_{X} X_{R}^{\dagger} \exp\left( \int dz A_5(z) \right) X_L \,+\, {\rm c.c.}
\eeq
is gauge-invariant, and, if $M_X$ is large enough (of order a few hundred GeV) can render $X$ unobservable in the Tevatron and LEP-2 searches. The Chern-Simons term needed to restore the gauge invariance on the IR and UV branes has the form (in the electroweak sector)
\beqa
&S_{CS5}=\frac{1}{24\pi}\int d^4x \,\int dz \,\Bigl[ N_c \,N_L \,N_R \,x^3 \,\,dBBdB\,\nn\\
&+ \,3 N_c \, N_R \, x \, \text{Tr}_{2L} \left[ dA_{L}dA_{L} \right]B \,+\,
3 N_c \, N_L \, x\, \text{Tr}_{2R} \left[ dA_{R} dA_{R} \right]B \Bigr] \,.
\eeqa
The corresponding WZ terms have the form
\beqa
&\Gamma_{WZ}=-\frac{1}{24\pi^{2}}\int d^4x \,\epsilon^{\mu\nu\rho\sigma} \frac{1}{4}N_{c}N_{L}N_{R}x^{3}\left[B_{\mu}(R^{\prime})B_{\nu}(R)-B_{\mu}(R)B_{\nu}(R^{\prime})\right] \times \left[B_{\rho\sigma}(R^{'})+B_{\rho\sigma}(R)\right]\nn\\
&-\frac{3}{24\pi^{2}}\int d^4x \, \epsilon^{\mu\nu\rho\sigma} N_{c}N_{L}x\frac{1}{4}\text{Tr}_{2R}\{\,\big[\,B_{\mu}(R^{\prime})A_{R{\nu}}(R)-B_{\mu}(R)A_{R{\nu}}(R^{\prime})-A_{R{\mu}}(R)B_{\nu}(R^{'})+A_{R{\mu}}(R^{'})B_{\nu}(R)\,\big]\nn\\
&\times \left[F_{R\rho\sigma}(R)+F_{R\rho\sigma}(R^{'})\right]\,\}\nn\\
&-\frac{3}{24\pi^{2}}\int d^4x \, \epsilon^{\mu\nu\rho\sigma} N_{c}N_{R}x\frac{1}{4}\text{Tr}_{2L}\{\,\left[\,B_{\mu}(R^{\prime})A_{L{\nu}}(R)-B_{\mu}(R)A_{L{\nu}}(R^{\prime})-A_{L{\mu}}(R)B_{\nu}(R^{'})+A_{L{\mu}}(R^{'})B_{\nu}(R)\,\right]\nn\\
&\times \left[F_{L\rho\sigma}(R)+F_{L\rho\sigma}(R^{'})\right]\,\}\,.
\eeqa
In general, depending on the ultraviolet completion of the Higgsless model, a number of $X$ fields, in different gauge representations, may be present; one would then need to add the corresponding CS terms. Since the space of possibilities is rather large, we will study one particularly simple representative example in this paper, which should be sufficient to get a sense of the LHC sensitivity to the topological terms. We assume $N_X$ right-handed fields on the IR brane, with gauge charges
\beq
X_i \in ({\bf 3}, {\bf 1}, {\bf 2}, 1/6),~~~i=1\ldots N_X\,,
\eeq
and $N_X$ left-handed fields, in conjugate representations, on the UV brane. The form of the CS and WZ terms contributed by these fields is exactly the same as in the minimal model, Eqs.~\leqn{CS} and~\leqn{WZ}, but with the overall coefficient $N_X$ multiplying both terms. This form will allow us to simply rescale the results of the analysis in the minimal model. We defer a more comprehensive study of the LHC sensitivity to topological terms in models with arbitrary fermion content to future work.

\section{Conventional and Chern-Simons Induced $Z^\prime$ Decays}
\label{sec:decays}

While vertices involving 2 electrically charged and 1 electrically neutral gauge boson already exist at tree level, there are no tree-level vertices involving 3 neutral gauge bosons, either Abelian or non-Abelian. Topological interactions, on the other hand, induce such vertices. The most obvious potentially observable consequence is new decay channels, of the form $V^i\to V^jV^k$, where all three $V$'s are neutral gauge bosons, and indices correspond to their KK numbers. The KK decomposition of the 5D neutral gauge fields is
\beqa
B_\mu(x,z) &=& \sum_{k=0}^\infty \psi^{B}_k(z)\,A^k_\mu(x)\,,\nn \\
A^3_R(x,y) &=& \sum_{k=0}^\infty \psi^{R}_k(z)\,A^k_\mu(x)\,, \label{KK}\\
A^3_L(x,y) &=& \sum_{k=0}^\infty \psi^{L}_k(z)\,A^k_\mu(x)\,\nn.
\eeqa
where $A^k$ are 4D gauge fields:  $A^0$ is the SM photon, $A^1$ is the SM $Z^0$ boson, and $A^k$ with $k\geq 2$ are non-SM, heavy states. The wavefunctions $\psi^{(X)}_k(z)$, along with the 4D masses of the fields $A^k$, are obtained by solving the 5D equations of motion with appropriate boundary conditions~\cite{HL}.  (A sample mass spectrum of the first few KK modes is given in Table~\ref{tab:KKparams}.) In particular, the photon wavefunctions are flat, $\psi^{(X)}_0(z)=$const, while all other fields have non-trivial profiles.

Substituting Eq.~\leqn{KK} into Eqs.~\leqn{CS} and~\leqn{WZ}, and performing the integration over $z$ in the former, yields a set of 4D ``topological" vertices of the form
\beq
\tilde{\kappa}_{ijn}\epsilon^{\mu\nu\rho\sigma}A^i_\mu A^j_\nu \partial_{\rho}A^n_\sigma\,.
\eeq
Decomposing $\tilde{\kappa}_{ijn}=\tilde{\kappa}_{(ij)n}+\tilde{\kappa}_{[ij]n}$, it is clear that only the antisymmetric part contributes; we will therefore define $\kappa_{ijn}=\tilde{\kappa}_{[ij]n}$. The contribution to these couplings from the Chern-Simons terms is
\beqa
&\kappa^{{\rm CS}}_{ijn} = \frac{1}{24\pi^2}\,\int dz \left[ \frac{1}{36}\left( \frac{d\psi_i^B}{dz}\,\psi^B_j -\psi_i^B\,\frac{d\psi_j^B}{dz} \right) \psi_n^B + \frac{1}{4}\left(\frac{d\psi_{i}^{B}}{dz}\psi_{j}^{R}-\psi_{i}^{R}\frac{d\psi_{j}^{B}}{dz} - \psi_{i}^{B}\frac{d\psi_{j}^{R}}{dz} + \frac{d\psi_{i}^{R}}{dz}\psi_{j}^{B}\right) \psi_n^R \right]\,,
\eeqa
while the contribution from the WZ effective action is
\beqa
& \kappa^{{\rm WZ}}_{ijn} \,=\, -\frac{1}{24\pi^2}\,\Bigl[ \frac{1}{72}\,\psi^B_i(R^\prime) \psi_j^B(R) \left( \psi_n^B(R)+\psi_n^B(R^\prime)\right) \nn \\
& + \frac{1}{8}\left( \psi_i^B (R^\prime) \psi_j^R(R) - \psi_i^B(R) \psi_j^R(R^\prime) \right)\left(  \psi_n^R(R^\prime)+ \psi_n^R(R)\right)\Bigr]\,-(i\leftrightarrow j)\,.
\eeqa
The total topological coupling is
\beq
\kappa_{ijn} \,=\, \kappa^{{\rm CS}}_{ijn} + \kappa^{{\rm WZ}}_{ijn} \,.
\eeq
For any $i$ and $n$, $\kappa_{iin}=0$ by symmetry. In particular,  $\kappa_{000}=0$, that is, there is no three-photon vertex. In addition, all vertices involving two photons and an arbitrary massive boson vanish: $\kappa_{00n}$ due to symmetry and $\kappa_{n00}$ due to a cancellation between the CS and WZ terms. This is guaranteed by gauge invariance, and implies the absence of anomalous photon pairs, {\it e.g.} from $Z^0\to\gamma\gamma$.  However, there are no such cancellations in $\kappa_{n10}$ and $\kappa_{n11}$ for $n\geq 2$, which induce the KK-mode decays
\beq
A^n \to Z^0\gamma\,,~~~A^n\to Z^0Z^0\,~~(n\geq 2)\,,
\label{decays}
\eeq
respectively. These decays, which do not occur at tree level, would provide a clear signature for the topological interactions.

\begin{table}[t]
\begin{center}
\begin{tabular}{|c|c|}
  \hline
  State & $M$ (GeV) \\
  \hline
  $A^2$ & $694.1$ \\
  $A^3$& $717.6$  \\
  $A^4$ & $1111$ \\
  $A^5$ & $1577$ \\
  \hline
  \hline
\end{tabular}\qquad \begin{tabular}{|c|c|c|}
  \hline
   & $A^2$ & $A^3$ \\
  \hline
 $\bar{\nu}_{L}\nu_{L}$       & $0.262\,g_{Z_{0}\bar{\nu}_{L}\nu_{L}}$ &  $0.169\,g_{Z_{0}\bar{\nu}_{L}\nu_{L}}$\\
  $\bar{\nu}_{R}\nu_{R}$       & $0.016\,g_{Z_{0}\bar{\nu}_{L}\nu_{L}}$ & $0.017\,g_{Z_{0}\bar{\nu}_{L}\nu_{L}}$ \\
  $\ell^+_{L}\ell^-_{L}$     & $0.098\,g_{Z_{0}\bar{\ell}_{L}\ell_{L}}$ &   $0.166\,g_{Z_{0}\bar{\ell}_{L}\ell_{L}}$\\
  $\ell^+_{R}\ell^-_{R}$     & $0.199\,g_{Z_{0}\bar{\ell}_{R}\ell_{R}}$ & $0.258\,g_{Z_{0}\bar{\ell}_{R}\ell_{R}}$ \\
  \hline
  $\bar{u}_{L}u_{L}$           & $0.073\,g_{Z_{0}\bar{u}_{L}u_{L}}$ &  $0.167\,g_{Z_{0}\bar{u}_{L}u_{L}}$\\
  $\bar{u}_{R}u_{R}$           & $0.290\,g_{Z_{0}\bar{u}_{R}u_{R}}$ & $0.389\,g_{Z_{0}\bar{u}_{R}u_{R}}$ \\
  $\bar{d}_{L}d_{L}$           & $0.184\,g_{Z_{0}\bar{d}_{L}d_{L}}$ & $0.168\,g_{Z_{0}\bar{d}_{L}d_{L}}$ \\
  $\bar{d}_{R}d_{R}$           & $0.515\,g_{Z_{0}\bar{d}_{R}d_{R}}$ & $0.600\,g_{Z_{0}\bar{d}_{R}d_{R}}$  \\
  \hline
  $\bar{t}_{L}t_{L}$            & $1.805\,g_{Z_{0}\bar{t}_{L}t_{L}}$ & $2.748\,g_{Z_{0}\bar{t}_{L}t_{L}}$  \\
  $\bar{t}_{R}t_{R}$            & $0.661\,g_{Z_{0}\bar{t}_{R}t_{R}}$ & $0.956\,g_{Z_{0}\bar{t}_{R}t_{R}}$ \\
  $\bar{b}_{L}b_{L}$            & $1.005\,g_{Z_{0}\bar{b}_{L}b_{L}}$ &  $1.070\,g_{Z_{0}\bar{b}_{L}b_{L}}$ \\
  $\bar{b}_{R}b_{R}$            & $0.224\,g_{Z_{0}\bar{b}_{R}b_{R}}$ & $0.281\,g_{Z_{0}\bar{b}_{R}b_{R}}$\\
  \hline
  $W^+W^-$    & $0.059\,gc_{W}$ &  $0.074\,gc_{W}$  \\
  \hline
   \hline
\end{tabular}
\caption{\label{tab:KKparams} Masses and couplings of the low-lying neutral electroweak Kaluza-Klein gauge bosons in the Higgsless model with the following parameters:
$1/R=10^{8}$ GeV, $1/R^\prime=283.27$ GeV, $m_t=172.5$ GeV. For the 1st and 2nd generation quarks and all leptons, $c_L=-c_R=0.46$; for the 3rd generation quarks, $c_L=0.46$, $c_R=-0.05$~\cite{HL_S}.}
\end{center}
\end{table}

The matrix element for decay $A^n\to Z^0Z^0$ has the form
\beq
{\cal M}\left(A^n(p)\to Z^0(k)Z^0(k^\prime)\right)\,=\,2\,\kappa_{n11}\,(k^\prime-k)_\rho\,
\epsilon^{\mu\nu\rho\sigma}\,\vareps^n_\mu(p)\,\vareps^1_\nu(k)\,\vareps^1_\sigma(k^\prime)\,,
\eeq
while the matrix element for the decay $A^n\to Z^0\gamma$ is
\beq
{\cal M}\left(A^n(p)\to Z^0(k)\gamma(k^\prime)\right)\,=\,2\,\left[ (\kappa_{0n1}-\kappa_{10n}) k_\rho - (\kappa_{10n}-\kappa_{n10}) k^\prime_\rho \right]\,\epsilon^{\mu\nu\rho\sigma}\,\vareps^n_\mu(p)\,\vareps^1_\nu(k) \vareps^0_\sigma(k^\prime)\,.
\eeq
The corresponding decay widths are
\beqa
\Gamma(A^n\to Z^0Z^0) &=& \frac{\kappa_{n11}^2}{24\pi}\,\frac{m_n^3}{m_Z^2} \,\left( 1-4x \right)^{5/2}\,,\nn \\
\Gamma(A^n\to Z^0 \gamma) &=& \frac{\kappa_2^2}{24\pi}\,\frac{m_n^3}{m_Z^2}\,(1-x)\,\Bigl[ 1 \,+\,
\frac{\kappa_{1}^{2}-10\kappa_{1}\kappa_{2}-\kappa_{2}^{2}}{\kappa_2^2}\,x \,+\, \frac{10\kappa_1^2+8\kappa_1\kappa_2-\kappa_2^2}{\kappa_2^2}\,x^2 \nn \\ & &~~~\,+\,
\frac{(\kappa_1+\kappa_2)^2}{\kappa_2^2}\,x^3\,\Bigr]\,.
\eeqa
where $x=m_Z^2/m_n^2 \ll 1$, and in the second line we defined
\beq
\kappa_1 = \kappa_{0n1}-\kappa_{10n},~~~\kappa_2 = \kappa_{10n}-\kappa_{n10}\,.
\eeq
In our analysis, we will focus on the two lowest-lying non-SM KK modes, $A^2$ and $A^3$, since they have by far the largest production cross sections at the LHC. (These modes are closely degenerate in mass, as can be seen from Table~\ref{tab:KKparams}, while the next KK excitation is much heavier.)

\begin{table}[t]
\begin{center}
\begin{tabular}{|c|c|c|}
  \hline
  \hline
  Mode                   &    $\Gamma$ (GeV)     & Br$(A^2)$ \\
  \hline
  $t\bar{t}$   &    $5.58$            & 30.6\%\\
  $b\bar{b}$   &    $2.88$            & 15.8\%\\
  $u\bar{u}, c\bar{c}$   &   $0.08$   & 0.43\%\\
  $d\bar{d}, s\bar{s}$   &   $0.24$ & 1.3\%\\
  $\ell^+\ell^-$       &   $0.042$  & 0.23\%\\
  $\nu\bar{\nu}$         &   $0.27$  & 1.48\%\\
  $W^{+}W^{-}$   &   $8.93$ & 49.0\%\\
  \hline
  $ Z^{0}Z^{0}$           &   $0.14$   & 0.79\%\\
  $\gamma Z^{0}$         &   $0.06$   & 0.32\% \\
  \hline
  Total                &  $18.22$       &  100\%\\
  \hline
  \hline
\end{tabular}\quad \begin{tabular}{|c|c|c|}
  \hline
  \hline
 Mode                  &   $\Gamma$ (GeV)     & Br$(A^3)$ \\
  \hline
  $t\bar{t}$   &    $13.4$    & 39.4\%\\
  $b\bar{b}$   &    $3.4$     & 9.9\%\\
  $u\bar{u}, c\bar{c}$   &    $0.22$ & 0.65\%\\
  $d\bar{d}, s\bar{s}$   &    $0.23$  & 0.68\%\\
  $\ell^+\ell^-$ &  $0.090$ & 0.26\%\\
  $\nu\bar{\nu}$&   $0.12$  & 0.34\%\\
  $W^{+}W^{-}$&   $16.5$      & 48.5\%\\
  \hline
  $Z^{0}Z^{0}$        &   $0.06$     & 0.17\%\\
  $\gamma Z^{0}$      &   $0.05$     & 0.15\% \\
  \hline
 Total                &  $34.13 $     &  100\%\\
  \hline
  \hline
\end{tabular}
\caption{Partial decay widths and branching ratios of $A^2$ (left) and $A^3$ (right), including conventional
and topologically induced decays, for the same model parameters as in Table~\ref{tab:KKparams}. The
minimal Higgsless model (see Sec.~\ref{sec:min}) is assumed.}
\label{tab:widths}%
\end{center}
\end{table}

Tree-level decays of the neutral gauge KK modes are $A^{2,3}\to W^+W^-$ and $A^{2,3}\to f\bar{f}$, where $f$ are SM fermions. In the Higgsless model, the decays of the KK gauge bosons into the first- and second-generation quarks, as well as all leptons, are strongly suppressed; decays into the third-generation quarks, however, can be significant. The partial decay widths are given by
\beqa
\Gamma(A^n\to W^+W^-) &=& \frac{g_{n+-}^2}{48\pi}\,\frac{m_n^5}{m_W^4}\,\sqrt{1-4x_w}\,\left(
\frac{1}{4}+4x_w-17x_w^2-12x_w^3 \right)\,,\nn \\
\Gamma(A^n\to f\bar{f}) &=& \frac{m_n}{24\pi}\,\sqrt{1-4x_f}\,\Bigl[ (a^2+b^2)\,(1-x_f)+6abx_f \Bigr]\,,
\eeqa
where the coupling to the fermion of the form $\bar{f}A^\mu\gamma_\mu (aP_L+bP_R)f$ is assumed,
and we defined $x_w=m^2_W/m^2_n$, $x_f=m_f^2/m^2_n$.
The tree-level coupling constants $g_{n+-}$, $a$ and $b$ are obtained by integrating the products of appropriate 5D profiles over the $z$ coordinate. Numerical values of the relevant coupling constants, for one representative model parameter point, are listed in Table~\ref{tab:KKparams}. Notice the enhanced couplings to the 3rd generation right-handed quarks, due to their localization near the TeV brane where the KK gauge bosons are also localized.

The partial decay widths and branching ratios for $A^2$ and $A^3$, for a representative model parameter point in the {\it minimal} model, are shown in Table~\ref{tab:widths}. The branching ratio into the ``topological" decay modes in the minimal model is of order 1\% for $A^2$ and 0.3\% for $A^3$. While rare, these modes are not negligible, and thus could be observed with a sufficiently large sample of KK gauge bosons at the LHC. In the {\it non-minimal} model, the partial decay rates into the topological modes are
enhanced by a factor $(N_X+1)^2$, and the branching ratio grows rapidly with $N_X$. Another observation, important for the analysis below, is that the widths of $A^2$ and $A^3$ are comparable to or larger than their mass difference, so in practice the two resonances will not be resolved at the LHC. We will always combine the signal from these two KK states.

\section{Collider Phenomenology}
\label{sec:LHC}

\begin{figure}[t]
\bc
\includegraphics[scale=0.9]{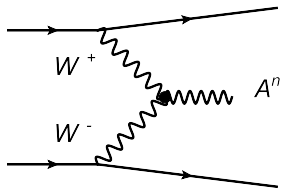}
 \ec
 \caption{\label{fig:VBF1} Production of $A^n$ gauge bosons in $pp$ collisions via the vector boson
 fusion process. (The additional contribution from $Z$ fusion through the topological vertex is highly suppressed and not shown here.)}
\end{figure}

Our goal is to evaluate whether the rare decays~\leqn{decays} can be observed at the LHC. There are two important mechanisms for producing the KK states $A^2$ and $A^3$: the conventional Drell-Yan (DY) process, $q\bar{q}\rightarrow A^2/A^3$, and the vector-boson fusion (VBF) process shown in Fig.~\ref{fig:VBF1}. The advantage of the VBF process is the high degree of model-independence~\cite{BMP}; in contrast, the DY process depends sensitively on the 5D profiles of light quarks, which dictate the quark couplings to $A^2$ and $A^3$. Precision electroweak constraints force the quark profiles to be delocalized, suppressing these couplings; nevertheless, due to the simple final state of the DY process, it may have a significant cross section. We consider both mechanisms in this section. The third potentially interesting production channel, associated production $pp\to A^{2,3}W$ and $pp\to A^{2,3}Z$, has a significantly lower cross section in the interesting parameter range, and will not be considered here.

\subsection{Vector Boson Fusion Production}

\begin{figure}[t]
\bc

%
%
%

\includegraphics[scale=1.1]{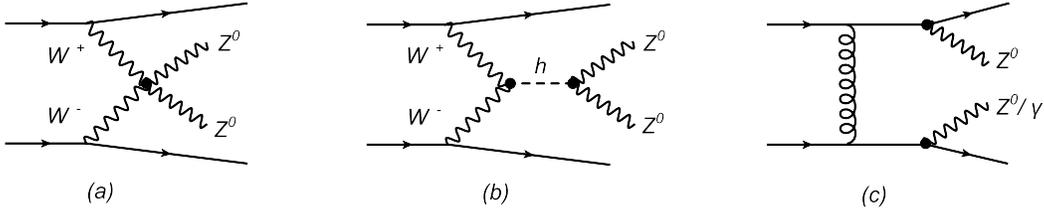}
\ec
\caption{\label{fig:VBF2}\label{fig:noVBF} Sample Feynman diagrams for the SM background processes $pp\to Z^0Z^0 qq$ (a,b,c), and $pp\to Z^0\gamma qq$ (c only).}
\end{figure}
\begin{figure}[th]
\centering
\includegraphics[scale=0.85]{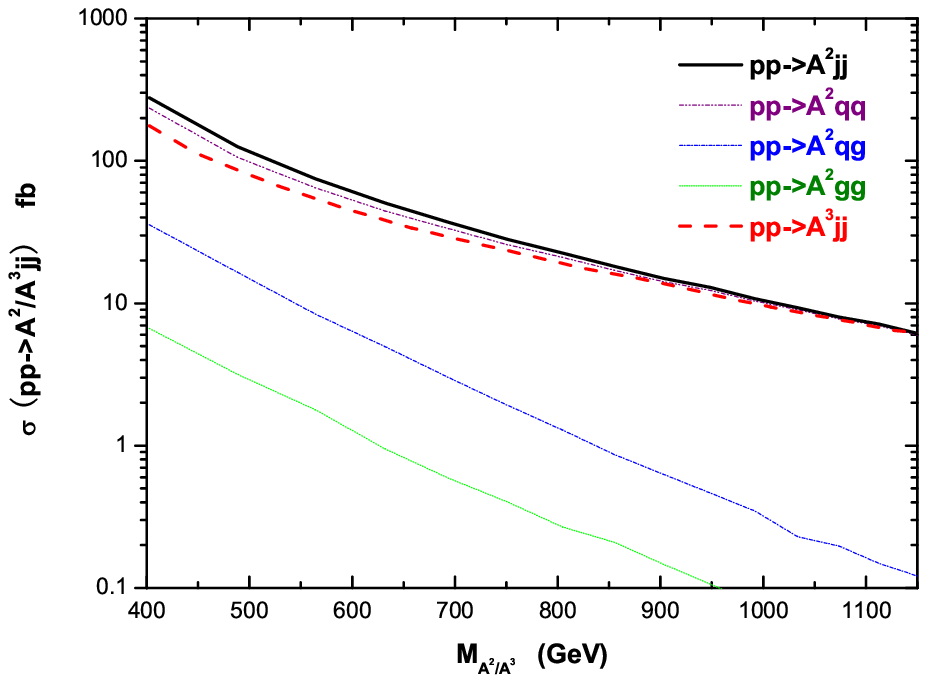}
\caption{Parton-level VBF production cross sections of the processes $pp\to A^{2,3}jj$ at the LHC ($\sqrt{s}=14$ TeV), calculated at tree level using CTEQ6L parton distribution functions~\cite{CTEQ}. Cuts in Eq.~\leqn{cuts1} are applied.}
 \label{fig:z1xsec}
\end{figure}

The VBF process in Fig.~\ref{fig:VBF1}, followed by decays~\leqn{decays}, results in the final states $Z^0Z^0jj$ and $Z^0\gamma jj$. In the SM, the final state $Z^0Z^0jj$ receives a contribution from VBF diagrams involving the 4-point $WWZZ$ vertex, as well as the Higgs production, see Fig.~\ref{fig:VBF2} (a,b). In addition, there is a contribution from non-VBF diagrams, such as the one shown in Fig.~\ref{fig:VBF2} (c). The latter contribution can be suppressed by demanding the jets to have large, opposite-sign rapidities, characteristic of the VBF kinematics. Specifically, we impose the cuts
\beq
2<|\eta(j_{1,2})|<4.5,~~E(j_{1,2})>300~{\rm GeV},~~p_{T}(j_{1,2})>20~{\rm GeV},~~\eta(j_{1})\eta(j_{2})<0\,.
\label{cuts1}
\eeq
For the $Z^0\gamma jj$ channel, there are {\it no} VBF-type diagrams in the SM, and the remaining non-VBF contributions are effectively suppressed by the same cuts. The total production cross sections of $A^2$ and $A^3$ at the LHC ($\sqrt{s}=14$ TeV), with cuts in Eq.~\leqn{cuts1}, are shown in Fig.~\ref{fig:z1xsec}. The cross sections were computed with a private Monte Carlo code, working at the leading order and using
CTEQ6L parton distribution functions~\cite{CTEQ}. In the mass range of interest, the cross sections are of order 10-100 fb, so that a substantial sample of KK gauge bosons can be produced with the expected 100-300 fb$^{-1}$ integrated luminosity.

\begin{figure}[t]
\begin{center}
\includegraphics[scale=0.6]{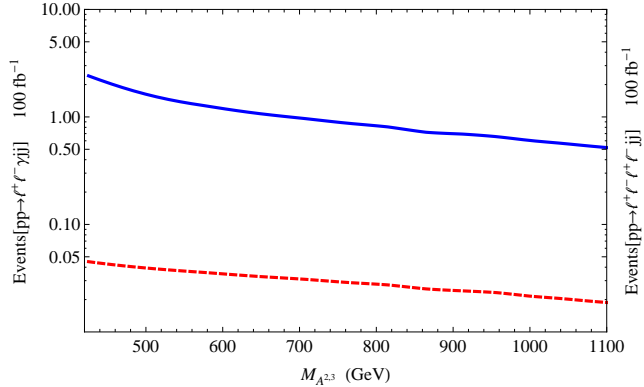}
 \caption{The expected number of  ``golden" events from topological decays of the KK gauge bosons
in a 100 fb$^{-1}$ VBF data sample at the LHC, in the minimal model, as a function of $M_{A^{2,3}}$ defined in Eq.~\leqn{Mave}. Blue/solid line: $\ell^+\ell^-\gamma+2j$ events; red/dashed line: $2\ell^++2\ell^-+2j$ events.}
 \end{center}
 \label{fig:rates}
\end{figure}

To avoid large background from QCD events, we focus on events with leptonic $Z^0$ decays, $Z^0\to \ell^+\ell^-$ with $\ell=e$ or $\mu$. This leads to the ``golden" final states $2\ell^++2\ell^-+2j$ and
$\ell^+\ell^-\gamma+2j$, where the jets are energetic and forward, passing the cuts~\leqn{cuts1}.
Taking into account the branching ratios, the expected number of events with the golden final states, produced via topological decays of the KK gauge bosons, is shown in Fig.~\ref{fig:rates}. It is clear that the $2\ell^++2\ell^-+2j$ final state is too rare to lead to an observable signature, even with very large integrated luminosity. We will therefore focus on the final state $\ell^+\ell^-\gamma+2j$, arising from topologically induced $Z\gamma$ decays of the KK gauge bosons.

The signal in the $\ell^+\ell^-\gamma+2j$ channel will appear as a Gaussian peak in the $s(\ell^+\ell^-\gamma)$ distribution, centered at the KK gauge boson mass, sitting on top of the SM background. (More precisely, in our case the signal appears as a double Gaussian peak, due to $A^2$ and $A^3$; however, the peaks effectively merge into one due to their widths.) In this situation, the background rate in the signal bin can in effect be measured from data by using the shoulder subtraction approach. We therefore estimate the expected significance $S$ of the signal by including only statistical uncertainty on the background; in other words, $S=N_{\rm sig}/\sqrt{N_{\rm bg}}$. We define the signal bin as the interval in $\sqrt{s(\ell^+\ell^-\gamma)}$ centered at
\beq
M_{A^{2,3}} \equiv \frac{M(A^2) + M(A^3)}{2}\,,
\label{Mave}
\eeq
with width $0.1\times M_{A^{2,3}}$. We compute the number of background events in this bin with a private Monte Carlo code which provides the leading-order cross section, using the CTEQ6L pdf's~\cite{CTEQ}. We impose additional cuts
\beq
|\eta(\ell)| < 2.5,~~~p_T(\gamma)>20~{\rm GeV}\,,
\eeq
to mimic the detector acceptance. (The efficiency of these cuts on the signal is close to one, since the signal $Z$ and $\gamma$ are central and energetic.) We estimate that the integrated luminosity needed to obtain $S=3$ ``observation" of the topologically induced rare decays at the LHC in this channel is of order $4-5$ ab$^{-1}$ throughout the interesting model parameter space. For $S=5$ ``discovery", an integrated luminosity of order
10 ab$^{-1}$ is required. Needless to say, such large data samples will be extremely difficult to obtain,
and are conceivably achievable only with a super-LHC luminosity upgrade of the original machine~\cite{super}.

\begin{figure}[t]
\bc
\includegraphics[scale=0.55]{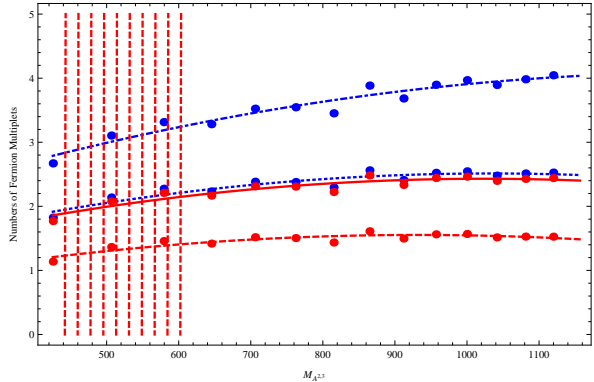}
\ec
\caption{\label{fig:Nreach}
 Number of heavy fermion fields required for $3\sigma$ observation and
 $5\sigma$ discovery of the topologically induced KK gauge boson decays in the
 non-minimal Higgsless model, as a function of $M_{A^{2,3}}$.
 The solid (red) curve: $S=5$ discovery for $300\,\rm{fb}^{-1}$;
 the dashed (red) curve: $S=3$ observation for $300\,\rm{fb}^{-1}$;
 the dash-dotted (blue) curve: $S=5$ for $100\,\rm{fb}^{-1}$;
 the dotted (blue) curve: $S=3$ for $100\,\rm{fb}^{-1}$.
 The shaded region is disfavored by precision electroweak tests~\cite{HL_S,Martin:2009gi}.}
\end{figure}

However, as discussed above, non-minimal models can produce significantly higher topological decay rates, resulting in a discovery with smaller integrated luminosity. This is illustrated in Fig.~\ref{fig:Nreach}, which shows the LHC reach in terms of $M_{A^{2,3}}$ and $N_X$, assuming realistic 100 fb$^{-1}$ and 300 fb$^{-1}$ data sets. It is clear that only a modest number of extra multiplets is required for discovery, and thus we encourage the experiments to take this possibility seriously in the LHC data analyses.

\subsection{Drell-Yan Production}

\begin{figure}[th]
\centering
\includegraphics[scale=0.85]{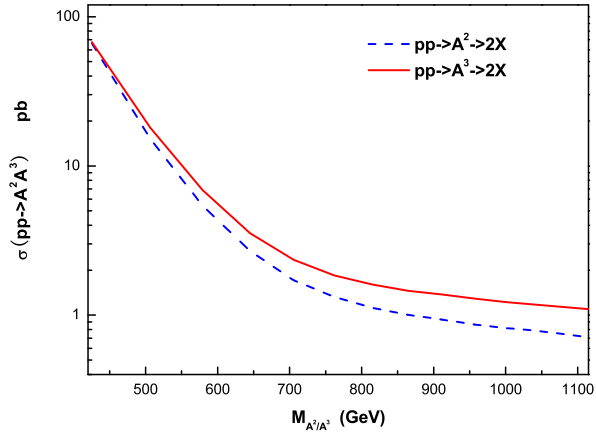}
\caption{Parton-level production cross sections of the DY processes $pp\to A^{2,3}$ at the LHC ($\sqrt{s}=14$ TeV), calculated at tree level using CTEQ6L parton distribution functions~\cite{CTEQ}. No cuts are applied.}
\label{fig:dyxsec}
\end{figure}
\begin{figure}[h]
\begin{center}
\includegraphics[scale=0.6]{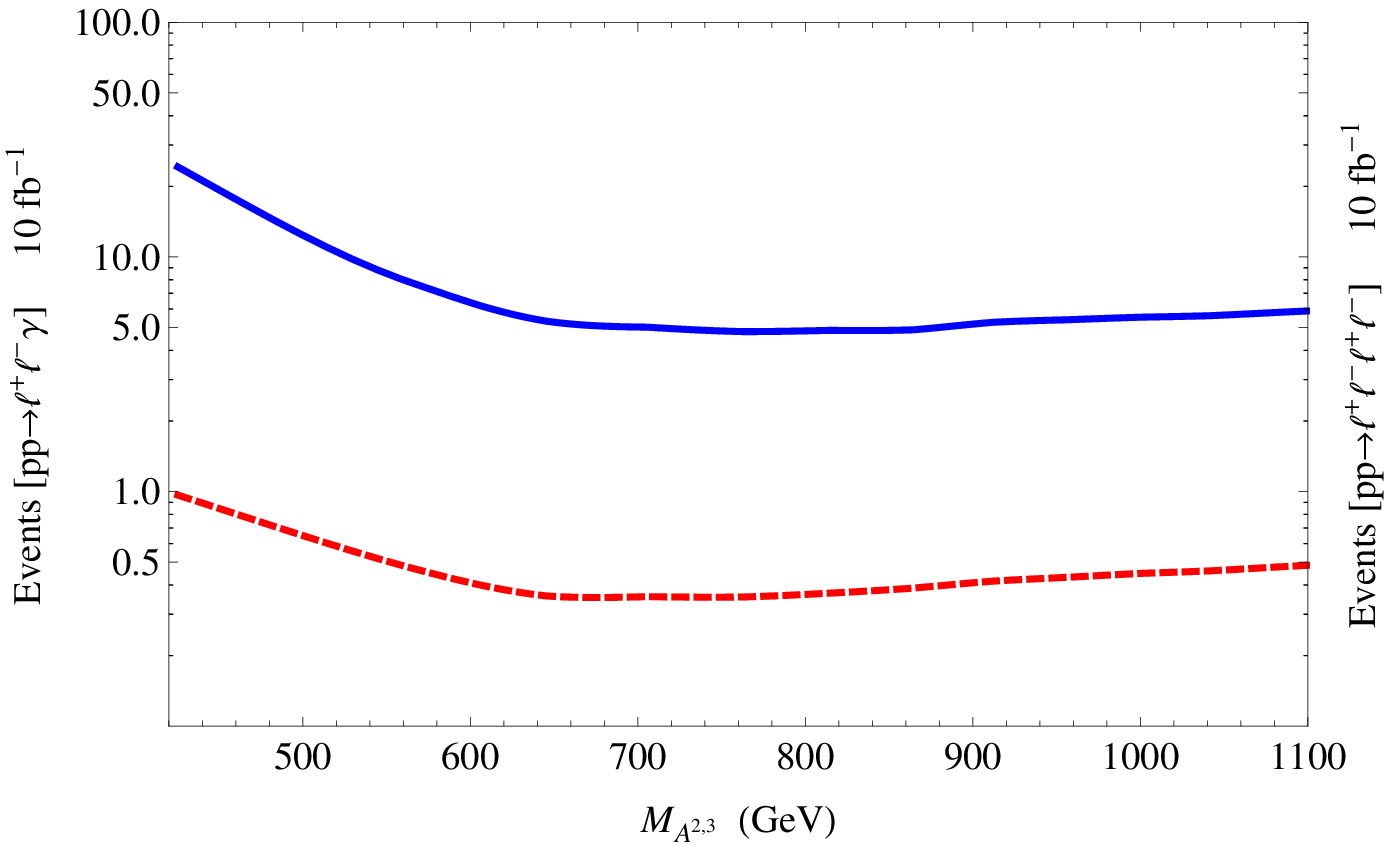}
 \caption{The expected number of  ``golden" events from topological decays of the KK gauge bosons
in a 10 fb$^{-1}$ DY data sample at the LHC, in the minimal model, as a function of $M_{A^{2,3}}$ defined in Eq.~\leqn{Mave}. Blue/solid line: $\ell^+\ell^-\gamma$ events; red/dashed line: $2\ell^++2\ell^-$ events.}
 \end{center}
\label{fig:dyrates}
\end{figure}

\begin{figure}[t]
\bc
\includegraphics[scale=0.55]{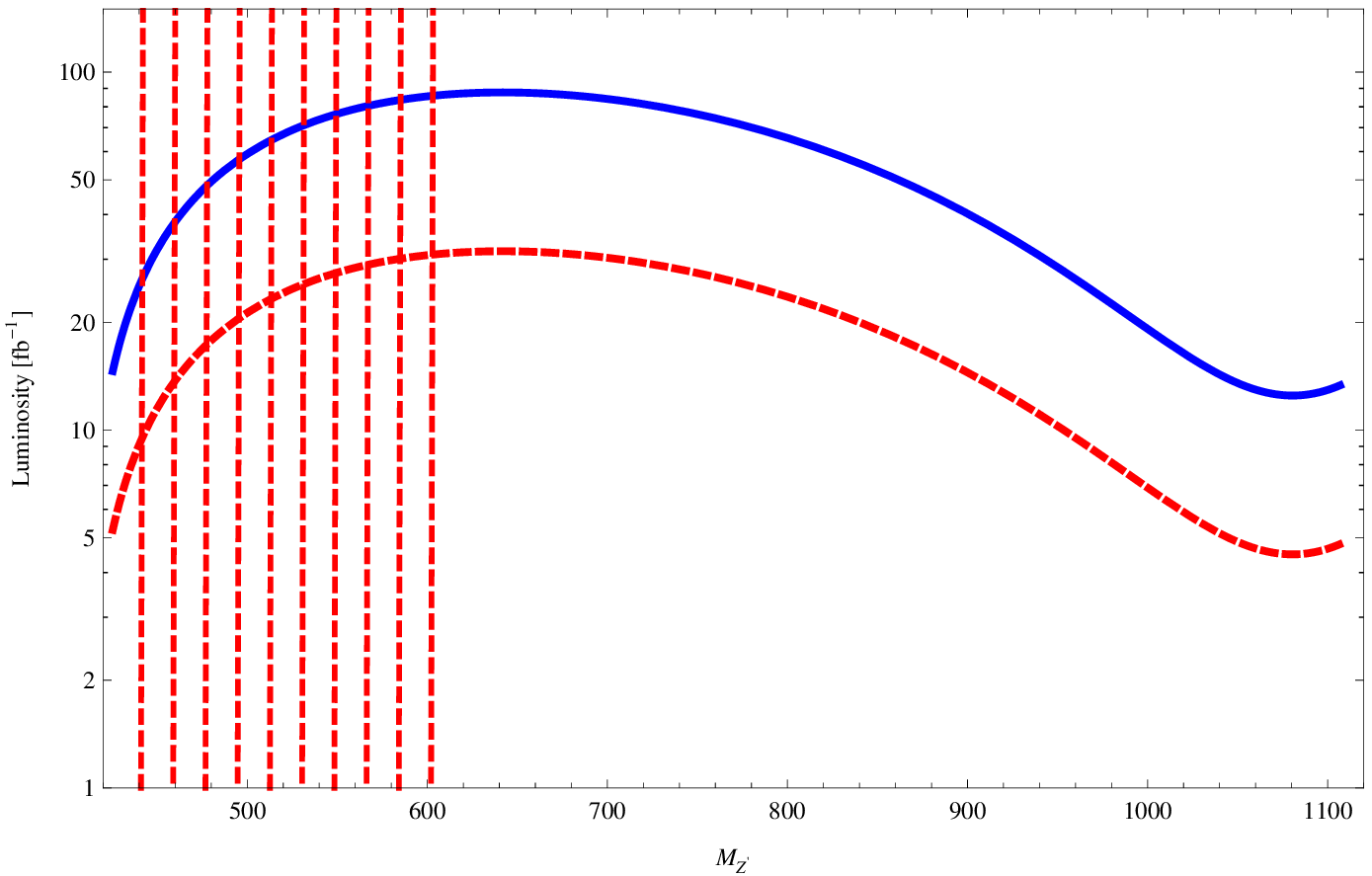}
\ec
\caption{\label{fig:dylum}
 Integrated luminosity at the LHC required for $3\sigma$ observation (dashed line) and
 $5\sigma$ discovery (solid line) of the topologically induced KK gauge boson decays
 in the Drell-Yan channel, as a function of $M_{A^{2,3}}$. Minimal Higgsless model is assumed. The shaded region is disfavored by precision electroweak tests~\cite{HL_S,Martin:2009gi}.}
\end{figure}

To study the Drell-Yan (DY) production of the KK states $A^2$ and $A^3$, we incorporated these particles, and their interactions listed in Table~\ref{tab:KKparams}, into the {\tt MadGraph/MadEvent} package~\cite{MG}. The total cross section of the DY process, $pp\to A^2/A^3$, is plotted in Fig.~\ref{fig:dyxsec}. It is clear from the figure that the cross section is quite large, one to two orders of magnitude larger than VBF. The number of events in the ``golden" final states produced by topological decays, $A^{2,3}\to Z^0\gamma\to \ell^+\ell^-\gamma$ and $A^{2,3}\to Z^0Z^0\to \ell^+\ell^-\ell^+\ell^-$, is plotted in Fig.~7.
Here we imposed the following acceptance cuts:
\beq
p_T(\ell^\pm) \geq 20~{\rm GeV,}~~~|\eta(\ell^\pm)| \leq 2.5.
\label{dycuts}
\eeq
Note that the number of events becomes approximately constant as the KK mass is increased, and in fact even increases slightly at large KK masses; this is due to the increasing branching ratio of the topological modes which offsets the decrease in the production cross section. Note also that the integrated luminosity assumed here is only 10 fb$^{-1}$, in contrast to 100 fb$^{-1}$ in the case of the VBF counterpart, Fig.~4. In addition to enhanced rates, the DY production channel followed by topological decays leads to a rather clean signature, since the SM backgrounds are small. We simulated the dominant backgrounds, $pp\to Z^0Z^0$ and $pp\to Z^0\gamma$, followed by leptonic decays of the $Z^0$, using {\tt MadGraph} and again imposing the cuts~\leqn{dycuts}.
The signal for topological decays of the KK bosons would again appear as a peak in the invariant mass distribution of the final-state particles. Using the same statistical procedure as for the VBF channel in the previous subsection, we estimate the integrated luminosity required for an observation and discovery of the topologically induced decays at the LHC, which is plotted in Fig.~\ref{fig:dylum}. Using the DY production channel, even the minimal model can be convincingly observed with realistic integrated luminosities: in fact 100 fb$^{-1}$ of data would be sufficient to discover the signal throughout the interesting mass range.
One needs to keep in mind that these results have been obtained with a particular choice of fermion bulk masses, see Table~\ref{tab:KKparams}, and may be quite sensitive to these parameters; on the other hand, the bulk masses themselves are quite tightly constrained by precision electroweak fits. It would be interesting to investigate the precise degree of model-dependence of the Drell-Yan process in this setup in the future.

\section{Conclusions}
\label{sec:concl}

Chern-Simons terms (and related global WZ interactions) arise naturally in 5D models of EWSB, whenever the fermion content is such that non-vanishing gauge anomalies are induced on each of the branes. If present, these topological terms lead to potentially observable consequences. A particularly interesting and clean signature is the decay of a KK-excited
gauge boson into a pair of SM neutral gauge bosons, $Z^0Z^0$ or $\gamma Z^0$, which does not occur at tree level. The prospects for observing such decays at the LHC depend on the KK gauge boson masses (which determine their production cross sections), and on the fermion content and localization in the model (which dictate the size of the topological couplings). In the ``cured" Higgsless model, where the first-level KK gauge bosons are predicted to be in the 600 GeV -- 1.2 TeV range, a significant sample of them would be produced at the LHC. We analyzed the prospects for discovering the anomalous decays induced by the topological terms in this model. Focusing on the highly model-independent vector boson fusion production channel for the KK gauge bosons,  ee found that in the minimal version of the model (with only the fermion fields required to reproduce the SM at low energies) the observation at the LHC is highly unlikely. However, if additional massive fermions are present at the TeV scale (which may be required in the ultraviolet completion of the theory), the signal rate may be greatly enhanced. We find that even a very modest number (two or three) of such extra fermions could be sufficient to render the signal in the $Z^0\gamma$ channel observable with about 100 fb$^{-1}$ of integrated luminosity at the LHC. We also analyze the LHC potential if the Drell-Yan production is used. While somewhat more model-dependent, these results are very promising: For the set of parameters adopted for our study, the LHC with 100 fb$^{-1}$ integrated luminosity would be able to discover the topological decays throughout the interesting KK gauge boson mass range, even in the minimal Higgsless model. It would be interesting to understand how sensitive this result is on the choice of fermion bulk masses, keeping in mind that these masses are already tightly constrained by precision electroweak fits.

This work can be extended in several directions. First, since the fermion zero modes in the Higgsless model are in fact {\it not} sharply localized close to the branes, it would be interesting to develop a formalism for calculating the CS terms without the brane-fermion approximation that we used. Second, it would be interesting to compute the CS terms in the fully realistic Higgsless model with the third-generation fermions embedded in extended representations to provide custodial protection of the $Zb\bar{b}$ couplings~\cite{HL_cust}. Finally, while we presented the general formulas for the CS and WZ terms in the presence of a massive fermion in an arbitrary gauge representation, we only explored phenomenological consequences of such fermions in one specific representation. It would be interesting to extend the phenomenological analysis to the more general case.

\vskip1cm

{\bf Acknowledgements}

We thank Csaba Csaki, David Curtin, Yuval Grossman, Johannes Heinonen, Jay Hubisz, Yu-Hsin Tsai, Henry Tye, Jiajun Xu and Yang Zhang for valuable discussions. We are grateful to Jing Shu for bringing Ref.~\cite{ref10} to our attention. This work is supported by the
U.S. National Science Foundation through grant PHY-0757868 and CAREER award PHY-0844667.
YHQ is supported in part by the Chinese Scholarship Council (No.~2008621151).

\bibliography{000}

\end{document}